\documentclass{kluwer}    
\usepackage{graphicx}
\def\corot{CoRoT}
\newdisplay{guess}{Conjecture}
\begin{document}                                                                                   
\begin{article}
\begin{opening}         
\title{The impact of CoRoT\ on close binary research} 
\author{Carla \surname{Maceroni}\email{maceroni@oa-roma.inaf.it}}  
\institute{INAF - Osservatorio Astronomico di Roma, Monteporzio Catone, Italy}
\author{Ignasi \surname{Ribas}\email{iribas@ieec.uab.es}}  
\institute{CSIC - Institut d'Estudis Espacials de Catalunya, Barcelona, Spain}

\runningauthor{C. Maceroni \& I. Ribas}
\runningtitle{CoRoT and close binary research}
\date{September 28, 2005}

\begin{abstract}
The space experiment \corot\ will provide continuous monitoring and
high accuracy light curves of about sixty thousand stars.
Selected binary systems will be observed in the Additional Program frame
as targets of long and continuous pointed observations. Moreover, 
thousands of new  binaries  will certainly be detected and hundreds of them
will have  extremely accurate light curves.
This will allow studies of fine effects on the light curves, monitoring of stellar
activity and, in combination with ground-based observations, will provide 
exquisite determination of stellar  parameters. 
 Among the new discoveries of interesting systems of special value will be 
those of  low mass binaries.
\end{abstract}
\end{opening}           
\section{\corot\ in a nutshell} 
\corot\  (COnvection, ROtation and planetary Transits) is a french-led international 
``small" space mission whose launch is scheduled for
August 2006.   
The mission\footnote{complete information on the mission is available at {\sf http://corot.oamp.fr}}
 focuses on two parallel ``core programs", asteroseismology and extra-solar planet search,
both requiring high accuracy photometry and  continuous monitoring.

The \corot\ payload consists of a 28cm telescope (1200mm focal length  and field of view of 3.8
degrees) which will fly in a polar inertial orbit. Its focal plane hosts two pairs of
2k$\times$2k CCDs, one dedicated to asteroseismology, the other to planetary transit searches, both
have a field of view of   $2.64^\circ \times 1.32^\circ$. 

 The defocused seismology field (PSF of 450 pixel for a G2 star)  will observe a 
maximum of ten bright targets  per run (5.7$\leq$V$\leq$9.5) with 
typical time sampling of 32$^{\mathrm{s}}$. Simultaneously, the (focused) exo-planet field will
observe up to 12000 fainter targets (11.5$\leq$V$\leq$16) with  typical time sampling 
of 512$^\mathrm{s}$. Color information will be available for the exo-planet targets brighter than
V=15, thanks to a prism providing very low resolution spectra.  The expected accuracy of the
photometry is  a few parts per million (ppm) for asteroseismology and 100 ppm for planet search. 

  The nominal mission duration is 2.5 years. Each year  will be split in
two 150$^{\mathrm{d}}$--long ``Long Runs" (LR), and two/one 20/30$^{\mathrm{d}}$--long ``Short Runs" (SR).
Pointing is restricted to the \corot\ ``Eyes", i.e. two 10$^\circ$ radius circles
centered at $\delta=0$ and, respectively, $\alpha=6^{\mathrm{h}}50^{\mathrm{m}}$ for the ``Anticenter" direction
and $\alpha=18^{\mathrm{h}}50^{\mathrm{m}}$ for the opposite, ``Center", direction.
The LR fields (much smaller than the ``Eyes") are chosen according to the best compromise between the two core
programs, and are constrained to a short list of preselected seismology ``primary targets".  Short Runs can 
instead be anywhere inside the Eyes.

Science outside the two core programs is included in the mission as specific "Additional Programs" (APs),
which can apply for both LR and SR observations. In practice, most APs will be performed in the Exo-planet field, 
because of the much larger number of available windows. 
\section{\corot\ and close binaries}
The best \corot\ assets are the accurate and stable photometry and
the continuous monitoring, with an  estimated duty cycle of 94\%.
These will provide  close binary  light curves of exquisite 
accuracy,  excellent phase coverage  and extending over a long time baseline.

The high accuracy with allow a thorough  study of fine effects in
the light curves by direct determination of second order
effects such as limb darkening and gravity darkening.  The latter can be 
derived for non eclipsing system as well, as it will clearly show up in
the frequency domain.

\begin{figure}    
\centerline{\includegraphics[width=.5\textwidth]{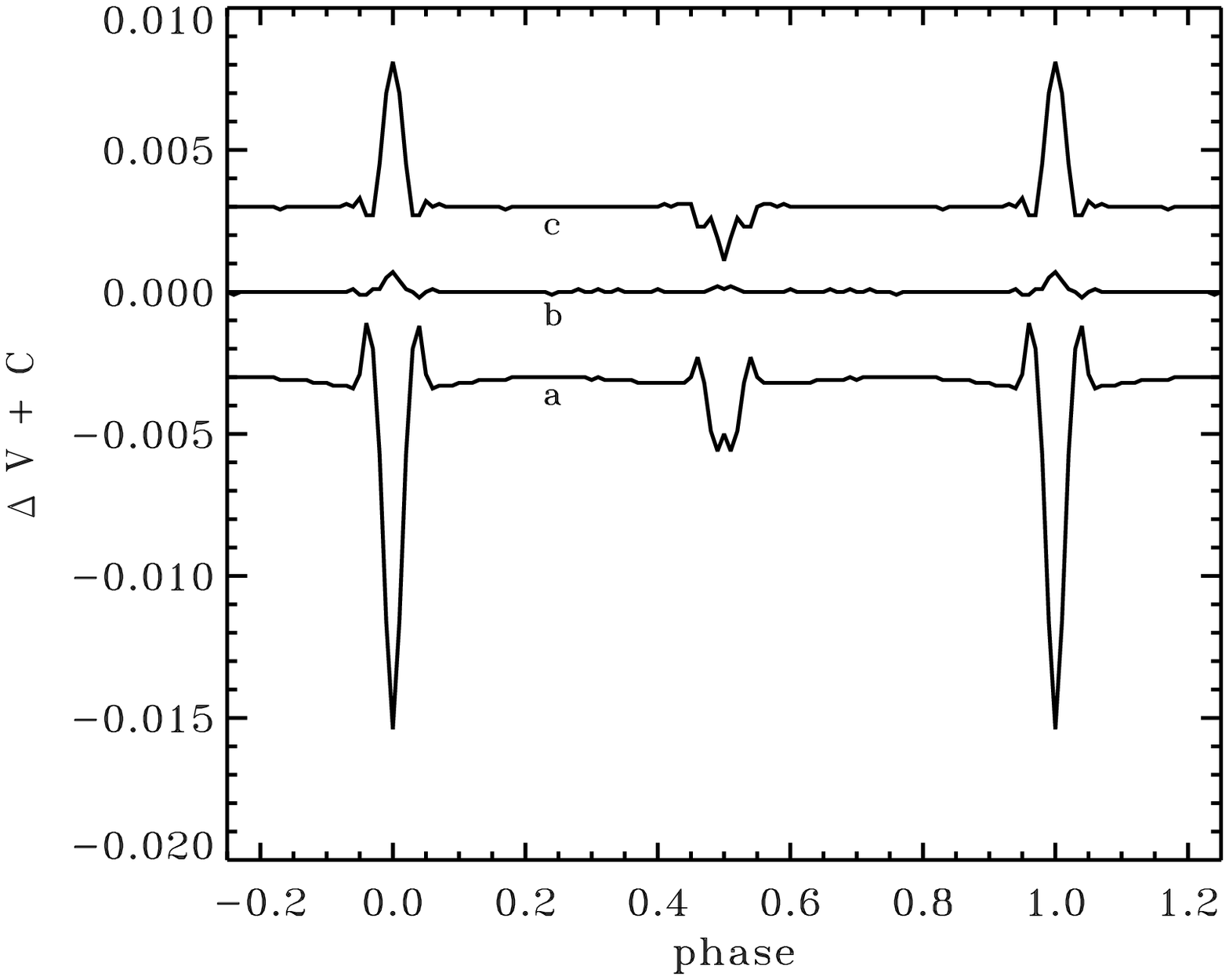}
\includegraphics[width=.5\textwidth]{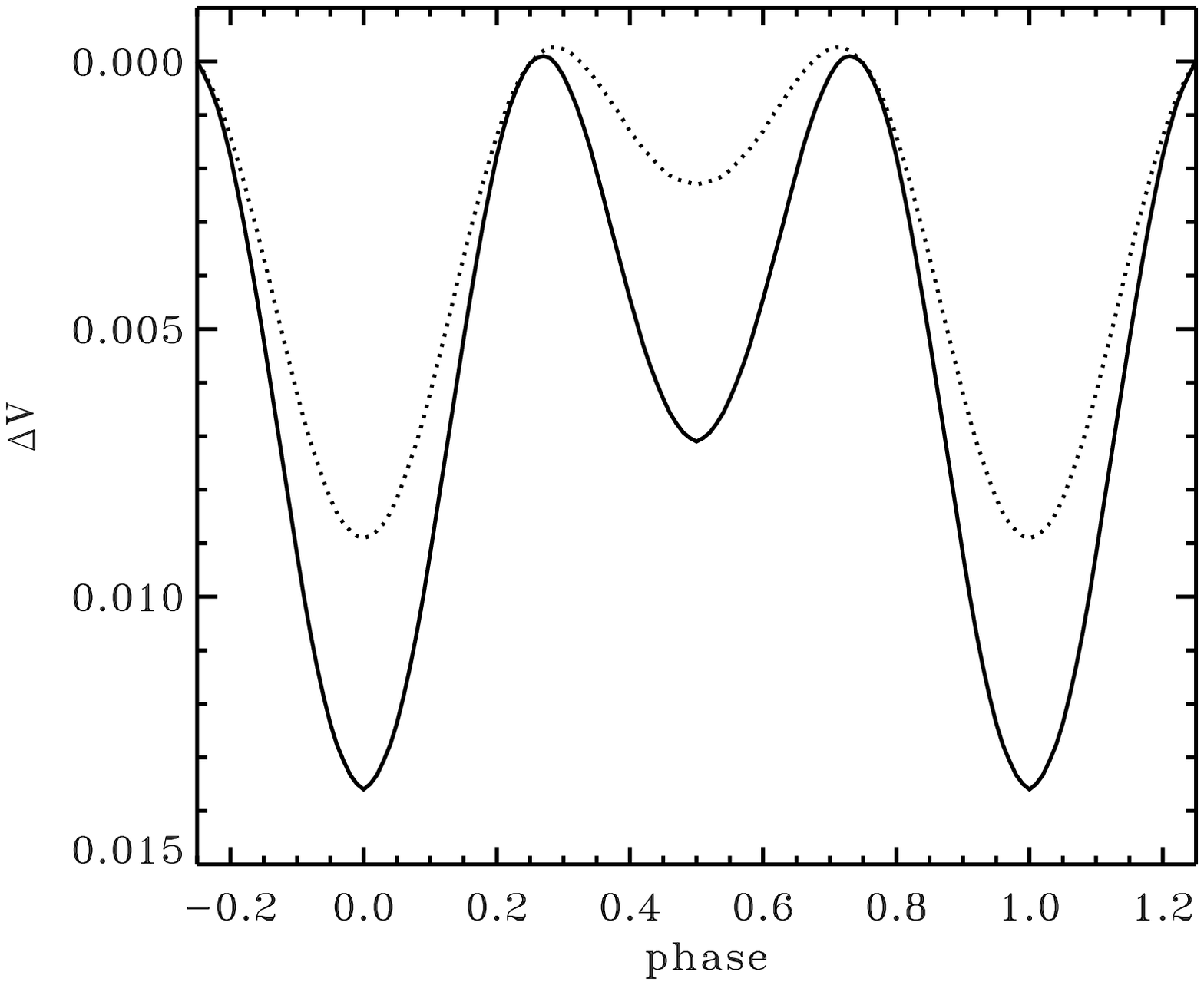}
}
\caption{Left panel: difference between synthetic light curve models with different assumptions on 
the limb darkening  law: a) linear (x=0.57) -- linear (x=0.67), b) linear (x=0.57) -- logarithmic, 
c) logarithmic law -- square root law. The parameters of the non-linear laws are taken from
Van Hamme (1993) for the corresponding component temperatures and gravities. For better readability  
curves a) and c) are shifted with respect to the zero line.
 Right panel: the effect of gravity darkening in a system with the same parameters as V805 Aql, 
but with $i=65^\circ$. The continuous line corresponds to a gravity darkening exponent $\beta=1.0$
(typical choice for radiative envelopes), the broken line to $\beta=0.32$ (convective envelopes).} 
\label{fig1} 
\end{figure}

The left panel of Fig.\ref{fig1} shows, as an example, the effect of different limb darkening laws 
on the light curve of a moderately close system. The synthetic light curves were computed
by the 2003 version of the Wilson and Devinney code \cite{wd71}.
For illustration purposes, we chose a model corresponding to the binary V805~Aql 
studied by \inlinecite{Popper84}; 
the main physical parameters of the model are: orbital period
P=2.41$^{\mathrm{d}}$, eccentricity $e=0$, inclination $i=86^\circ$, mass ratio $q=m_s/m_p=0.81$ (the index p stays for
primary and refers to the more massive component), fractional component radii
$r_p=0.18$ and $r_s=0.15$, effective temperatures $T_p=8184$ and  $T_s=7178$ K, 
fractional V-band luminosity $L_s/L_p=0.362$.
The figure shows the difference in magnitude between synthetic light curves obtained with 
different assumptions on limb darkening (different values of the coefficient for the linear
law, or different laws).  The difference between linear laws of the form: 
\begin{equation}
R_\lambda(\mu)=\frac{I_\lambda(\mu)}{I_\lambda(1)}= 1-x_\lambda(1-\mu),
\end{equation}
with $\mu=\cos\theta$, is obtained by changing the value of
the coefficient $x_\lambda$. 
Among the proposed  non-linear functional forms we considered a logarithmic relation:
\begin{equation}
R_\lambda(\mu)=1-x_\lambda(1-\mu) -y_\lambda \mu \ln \mu
\end{equation}
and a square root one:
\begin{equation}
R_\lambda(\mu)=1-x_\lambda(1-\mu) -y_\lambda (1 - \sqrt{\mu})
\end{equation}
see \inlinecite{vh93} and references therein.
The coefficients used to compute the differences (b) and (c) in Fig. \ref{fig1} have be taken, 
according to the component temperatures and gravities, from the 
tables by \inlinecite{vh93} (i.e. the differences are  between two
fits of the same atmosphere). 
For this reason the deviations are smaller than in the case of linear laws.
 At any rate the photometry by \corot\,, with its 10$^{-3}$--10$^{-4}$ mag 
accuracy, should allow to distinguish at least between linear and non-linear limb
darkening prescriptions. The directly determined limb darkening  can then be used to
constrain  model atmosphere details. 

The right panel of Fig.\ref{fig1} shows the amplitude of the ellipsoidal variation in
a system differing from  V805 Aql only for a lower value of the inclination (to 
avoid eclipses), $i=65^{\circ}$. In spite  of the small deviation from spherical
symmetry (the maximum difference in radius for the primary component is $<2$\%) 
the effect is large enough to be easily  measured by \corot. The determination of
gravity brightening and,  with the help of follow-up observations, of the related 
system configuration,  will yield precious information on close binary tidal evolution.

 The long monitoring of the same field will allow to study on a long time baseline 
the manifestations of stellar activity such as spots, flares, stellar activity cycles. 
These will provide information on rotational  period and differential rotation (from spot migration) 
in late-type components. 

A very young field, asteroseismology in close binaries,  combining the information from 
asteroseismology with that from eclipses will certainly profit of \corot\ assets. 
\begin{figure}
\centerline{\includegraphics[width=\textwidth]{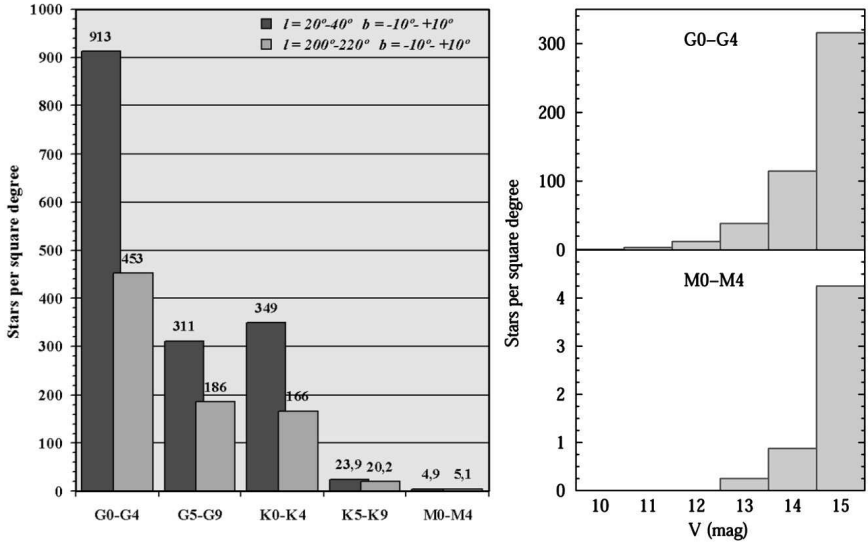}
}
\caption{Left panel: the expected number of G-M stars of luminosity class IV-V, in a 16$^\Box$ field 
inside the \corot\ eyes. Right panel: the distribution in apparent magnitude in the winter (AC)
field for G0-G4 and M0-M4 spectral type intervals.} 
\label{fig2} 
\end{figure}
Looking for (solar-type) oscillations in suitable eclipsing binaries would have the advantage of 
knowing the masses, the radii and the orbit inclination angle, and therefore the rotational velocity.
While binarity might increase the complexity of analysis, the identification of even non-radial 
pulsation modes is possible, as recently shown by several authors \cite{mkr05,gam05,esc05}. 
In particular eclipsing systems, with their near equator-on view, favor the detection
of sectorial modes (i.e modes with $\ell = |m|$, where nodal lines form lines of longitude).

The \corot\ eyes contain $\sim 200$ known system, but only a tiny fraction of them will
actually be observed during a Long Run 
\footnote{see the  link `Information' of the \corot\ 
{\it Thematic Team on binaries\ } web-page, at  {\sf http://thor.ieec.uab.es/binteam/}, for
details about the known binaries in the ``Eyes" and in the fields recently chosen
for the first two Long Runs.}. However, 
many interesting systems will certainly be discovered in the \corot\ fields.
Figure \ref{fig2} shows the results of a simulation with the Besan\c{c}on galaxy model
\cite{rob03}, the constraints are: spectral type G-M, luminosity class IV-V, 
V magnitude 5.5-15.5 (to include both sismo- and exo-fields). The results give
the number of star expected in a field of $4^\circ \times 4^\circ$ in the \corot\ 
Center and Anticenter directions.
 The magnitude distribution of the expected stars at the spectral extremes, and for
the winter (Anticenter) field, is also shown.

Photometric large surveys such as OGLE, MACHO, ASAS, STARE, Vulcan indicate that 0.5\%--1\%
of all monitored stars turn out to eclipsing binaries. 
From these percentages, and the density estimate, we obtained the maximum and minimum
number of eclipsing binaries for  an exo-planet field ($\sim 3.4^\Box$).
Table \ref{expbin} gives the results subdivided by spectral type intervals.
With a total number of five LRs and five/ten SRs some hundreds of new objects are expected only
along the MS.
Among those, the stars at the two extremes in mass  will be especially valuable
to better constrain the theoretical  models. In particular, at the low mass end of the
MS, the absolute elements obtainable by combining follow-up spectroscopy
with the \corot\ photometry will be precise enough to clarify what appears as
a serious discrepancy in the mass-radius relation between theoretical models and  
data from observed eclipsing binaries (see Ribas talk, these proceedings). 
\begin{table} %
\begin{tabular}{cc}                                        
\hline
Spectral type & expected EB number \\
\hline
G0-G4     & 7.2--31.1  \\
G5-G9     & 3.2--10.5  \\
K0-K4     & 2.9--11.9  \\
K5-K9     & 0.35--0.8  \\
M0-M4     & 0.09-0.17  \\
\hline
\end{tabular}
\caption[]{Expected binaries per 3.4$^\Box$ field}
\label{expbin}
\end{table}

In summary the ``small space experiment" \corot\ will provide great
opportunities not only in the field of asteroseismology and extra-solar planet
searches. Research on binary stars (and more generally on stellar physics)
will greatly benefit of its accurate photometry and stable monitoring
and many long standing questions will hopefully find a definite and clear answer.

\acknowledgements
We thank the members of the \corot\ ``Thematic Team" on binaries for useful
discussion  on the different aspects of binary research that can be addressed 
by the mission.
CM acknowledges the support from the MIUR-Cofin 2004 ``Asteroseismology".


\end{article}

\begin{thebibliography}{}
\bibitem[\protect\citeauthoryear{Escol{\`a}-Sirisi et al.} 
{2005}]{esc05} Escol{\`a}-Sirisi, E., Juan-Sams{\'o}, J., 
Vidal-S{\'a}inz, J., Lampens, P., Garc{\'{\i}}a-Melendo, E., 
G{\'o}mez-Forrellad, J.~M., and P. Wils: 2005, \newblock{`Detection of a 
classical $\delta$ Scuti star in the new eclipsing binary system HIP 
7666'},\newblock{\it Astronomy and Astrophysics\ }{\bf 434}, pp.~1063--1068 

\bibitem[\protect\citeauthoryear{Gamarova et al.} {2005}]{gam05} 
Gamarova, A.~Yu., Mkrtichian, D.~E. and E. Rodr{\'{\i}}guez: 2005, 
\newblock{`Mode identification in the ecplising binary systems with primary 
pulsating components'},
\newblock{\it ASP Conf. Ser. 333: Tidal Evolution 
and Oscillations in Binary Stars\ }{\bf 333}, pp.~258

\bibitem[\protect\citeauthoryear{Mkrtichian et al.} {2005}]{mkr05} 
Mkrtichian, D.~E., Rodr{\'{\i}}guez, E., Olson, E.~C., Kusakin, A.~V., Kim, 
S.-L., Lehmann, H., Gamarova, A.~Yu., and Y.~W. Kang : 2005, 
\newblock{`Pulsations in eclipsing binaries'},
\newblock{\it ASP Conf. Ser. 333: Tidal Evolution and Oscillations 
in Binary Stars\ }{\bf 333}, pp.~197

\bibitem[\protect\citeauthoryear{Popper}{1984}]{Popper84} 
Popper, D.~M.: 1984, 
\newblock{`Error analysis of light curves of detached eclipsing binary systems'},
\newblock{\it Astronomical Journal\ }, {\bf 89}, pp.~132--144

\bibitem[\protect\citeauthoryear{Robin et al.} {2003}]{rob03} Robin, 
A.~C., Reyl{\'e}, C., Derri{\`e}re, S., and S. Picaud: 2003, \newblock{`A 
synthetic view on structure and evolution of the Milky Way'},\newblock{\it 
Astronomy and Astrophysics\ }{\bf 409}, pp.~523--540 
 
\bibitem[\protect\citeauthoryear{Van Hamme}{1993}]{vh93} 
Van Hamme, W.: 1993,
\newblock{`New limb-darkening coefficients for modeling binary star light curves'},
\newblock{\it Astronomical Journal\ } {\bf 106}, pp.~2096--2117 %


\bibitem[\protect\citeauthoryear{Wilson and Devinney} {1971}]{wd71} 
Wilson, R. E.~and E.J. Devinney: 1971, \newblock{`Realization of 
Accurate Close-Binary Light Curves: Application to MR Cygni'},\newblock{\it 
Astrophysical Journal\ }{\bf 166}, pp.~605

\end{thebibliography}
\end{document}